# Exploring Application Logs

Janusz Sosnowski

Institute of Computer Science
Warsaw University of Technology, Warsaw, Poland
Email: jss@ii.pw.edu.pl

*Abstract*—This paper deals with the problem of analyzing application event logs in relevance to dependability evaluation. We present the significance of application logs as a valuable source of information on operational profiles, anomalies and errors. They can enhance classical approaches based on monitoring system logs and performance variables.

*Keywords; event monitoring, operational profiles, anomalies*

## I. INTRODUCTION

On-line monitoring of system operation is widely used technique in dependability to detect, handle and predict various problems (errors, anomalies). The monitoring processes can be performed at various levels (microarchitecture, architecture, system, application). System level monitoring gained big interest and significant practical results have been published ([1,2,4] and references). Application level monitoring seems to be neglected. We faced this problem in various projects and gained some experience.

Commercial and open-source programs quite often generate event logs related to their operation. In practice they differ significantly upon application, moreover the interpretation of registered events is not clear or even ambiguous. In Windows these events are stored either in a general application log or in dedicated logs. In Unix usually these events are stored in various files according to syslog logging scheme. Manual analysis of logs is impossible due to a large amount of generated events, the more that sometimes they are mixed within different log files. To get better knowledge on application log usefulness we have developed several tools facilitating tracing log morphology and mining their practical significance. In a large extent this is based on regular expressions and some data exploration algorithms.

In the sequel we give an outline of application log features, methodology of their analysis and some illustrative results.

## II. APPLICATION LOG FEATURES

General event logs (generated by the operating system) provide information on system operation and behaviour (logins of the users, system start-ups, closings, crashes, restarts, security problems, problems with I/O drivers, warnings, various information or notices, etc.). The scope of generated events depends upon the system load, configuration, etc. Recently operating systems include more and more application logs (it is visible for subsequent versions of Windows) which are generated by applications via operating system and stored in a general application log or directly stored in individual application logs. The registered events relate to the execution of the application. Having analysed a wide spectrum of application logs we have observed a big dispersion in formats and information contents. In a large extent this results from the fact that they are defined by the application designers. However some tries of standardization can be encountered e.g. in the case of SOA systems with logs characterising service and task executions.

Basing on the registered event logs we can evaluate system dependability, identify errors, anomalies, downtimes, etc. These problems may relate to hardware (permanent, intermittent, transient faults), software or maintenance problems. The space of possible problem sources is increasing and their identification becomes difficult due to high complexity of contemporary systems, frequent updates or upgrades of software, sporadic hardware enhancements or modifications, configuration changes, etc. Hence deeper analysis of event logs is required. To get better knowledge on this problem we have collected event logs (in particular application logs) from many computers (laptops, workstations, servers and a big cluster system). The collected data (within several data bases) has been explored using various tools developed in the Institute.

The crucial point is event categorization [3] as well as finding features of normal (operational profiles) or anomalous behaviour. Most application logs are characterised by free formats and text contents, so automatisation of their interpretation is a challenge. However in some systems a preliminary classification can be available. In the case of Windows systems many applications (e.g. MS Office, SQL) provide events with specified event source (e.g. application name), event ID, event type (error, warning, information) and a textual description of the event (message). It is worth noting that for the same event source, event ID and type there are possible different event specifications (message field). Moreover the same ID can appear for different event sources or types. Naturally all events comprise related to their occurrence time stamp. In the case of Unix (Linux) event logs we do not have event ID, so their categorization is more complex. Moreover upon the event severity (e.g. emergency, alert, critical, warning, debug, info) it can be loaded to appropriate log file via *syslog* program. Some application logs are very specific e.g. mail log generated by *sendmail* program comprises information on sender, receiver and servers involved in email transmission, mail identifier, mail size, number of receivers, transmission delays, used communication protocol, so the interpretation can be more sophisticated.

Analysing event logs we have concentrated on defining representative event classes which can be attributed to interesting situations, system behaviour, etc. Despite log diversity we can identify some specific field in event records e.g. time stamp, computer name, event source, ID (if available), etc. They can be useful in filtering data. Deeper analysis is needed within message texts, here we use regular expressions and some data exploration algorithms. Regular expressions are useful for selecting known events. Basing on some experience we can get a knowledge on events which correlate with problems of specific situations in the system. Moreover for some applications log diversity is low so we can even define all categories of mails. For *sendmail* program we have identified 53 event classes of the following types: 26 – info, 18, notice (e.g. invalid root address, user unknown, syntax error, exceeded message size) , 2 – debug, 2 – alerts , 1 - warning, , 4 – critical, 1 – security (the cardinality of different events within these classes was in the range from tenths to hundreds). In general direct (especially for complex applications) and complete event categorization with regular expressions is practically not possible. Hence we have developed data exploring algorithms which analyse event messages and identify constant (which occurred with high frequency) and variable texts (changing in the events). In this way we find so called variables (parameters) which are replaced with special symbols e.g. (…), some specific variables e.g. related to time stamp we can replace with wild card symbol (*). In this way we will get specifications of different event classes. For example for a cluster system which generated about 250 000 events (within 3 months) abstracting time stamp and PID reduced this set by 50%. Using the developed algorithm we have got about 12 000 event classes. Abstracting nodes this resulted in 5000 event classes. Further refinement resulted in about 300 classes.

In practice it is useful to combine regular expression approach with data exploration; moreover these processes can be done iteratively in several steps. In particular having identified variable fields we can match some regular expressions describing the structure of these fields and suggesting their meaning e.g. transmission port, web address, program version number, file name, error code. Deeper exploration can involve also some text mining (e.g. including text context). For this purpose we have developed some algorithms which identify word occurrences, specific phrases, etc. For example for application logs (collected for two years) of a frequently used laptop we have got over 42000 application event records (lines) comprising about 700 000 words (about 1500 different). The most frequent words were *error* (26000) and *update* (11 000). However *fault* and *failure* occurred 14 and 3 times, respectively. In the analysed system we had application logs in English and Polish, so this additionally complicated text mining. We are also interested in some phrases with *not* e.g. *not able*, *not capable*, which may relate to some problems. On the other hand the phrase *no user action is required* relates to non-critical events. Such statistics is useful in selecting keywords helpful in defining regular expressions. Looking for event classes we also compare text distance between messages in different event records.

Analysing operational profiles it is reasonable to find event profiles in particular sequences of events. We have performed complete analysis of pairs and some longer event sequences. For the considered laptop we have got 1307 different pairs of events. For 5% of pairs we have found that the first event always implies a specific successor event (confidence 1.0). In the case of 207 events the confidence ratio was c=0.5. For a standalone workstation about 9000 application events resulted in 42 pairs with c=0.5. This analysis proved high correlation of events, so checking these pairs in real system gives also useful information on unsuccessful program update of installation, frequency of correct updates, etc. We have also identified longer event sequences e.g. sequence of event IDs <900,1066,902,1003> occurred 300 times in the workstation (it relates to checking the originality of MS Office at every service start-up). Such characteristics describe individual computer profiles.

An important observation is that standard event types (error, info, etc.) are not sufficiently precise - coarse grained. Quite often event of type error does not relate to real error and on the opposite some info type events in fact relate to real errors (this problem has been observed also for syslog [4]). The developed data exploration identifies event classes which are more informative. Such classes can be inspected and refined by system administrators and users.

III. CONCLUSION

Application logs comprise valuable information on system operation. To derive this information we have developed some specialized data exploration procedures. Basing on our experience [1] we plan to integrate this with the analysis of other event (e.g. syslog, security) and performance logs.